\documentclass{epl}

\usepackage{graphicx,amsmath,amssymb}% Include figure files
\usepackage{dcolumn}% Align table columns on decimal point
\usepackage{bm}% bold math

\def\>{\rangle}
\def\<{\langle}

\title{Linear instability and statistical laws of physics}

\author{Giulio Casati\inst{1,2} \and Constantino Tsallis\inst{3,4} 
\and Fulvio Baldovin\inst{5,6}}

\institute{
\inst{1} Center for Nonlinear and Complex Systems - 
Universit\`a degli studi dell'Insubria, Via Valleggio, 11 - 22100 Como, Italy\\
\inst{2} CNR-INFM, Unit\`a di Como, and
Istituto Nazionale di Fisica Nucleare, sezione di Milano\\
\inst{3} Santa Fe Institute -
1399 Hyde Park Road,
Santa Fe, New Mexico 87501,  USA.\\
\inst{4} Centro Brasileiro de Pesquisas F\'\i sicas -
Rua Xavier Sigaud 150,
22290-180 Rio de Janeiro-RJ, Brazil.\\
\inst{5} INFM-Dipartimento di Fisica, Universit\`a di Padova -
Via Marzolo 8, I-35131 Padova, Italy\\
\inst{6} Sezione INFN, Universit\`a di Padova - Via Marzolo 8, I-35131 Padova, Italy
}%end of institute


\date{\today}

\begin{document}

\maketitle

\begin{abstract}
We show that a meaningful statistical description is possible in
conservative and mixing systems with zero Lyapunov exponent in
which the dynamical instability is only linear in time. More
specifically, (i) the sensitivity  to initial
  conditions is given by
  $ \xi =[1+(1-q)\lambda_q t]^{1/(1-q)}$
  with $q=0$; (ii) the statistical entropy
  $S_q=(1-\sum_i p_i^q)/(q-1) \;\;(S_1=-\sum_i p_i \ln p_i)$
  in the infinitely fine graining limit (i.e.,
  $W\equiv$ {\it number of cells into which the
  phase space has been partitioned} $\to\infty$),
  increases linearly with time only for $q=0$; (iii) a
  nontrivial, $q$-generalized, Pesin-like identity is
  satisfied,  namely the $\lim_{t \to \infty} \lim_{W
    \to \infty} S_0(t)/t=\max\{\lambda_0\}$. These facts
  (which are in analogy to
  the usual behaviour  of strongly
  chaotic systems with $q=1$), seem to open the door for a
  statistical description of conservative many-body
  nonlinear systems whose Lyapunov spectrum
  vanishes.

\end{abstract}

The possibility to derive macroscopic laws from the underlying microscopic deterministic dynamics
crucially depends on whether the latter satisfies or not the randomness assumptions
required by statistical mechanics.
In this respect exponentially unstable systems appear precisely those deterministic random systems tacitly required by macroscopic laws of physics.
In fact the exponentially unstable motion is called chaotic
since almost all orbits, though deterministic,
are unpredictable. This is a consequence of the Alekseev-Brudno theorem
\cite{alek} in the algorithmic theory of dynamical systems, according to which the
information $I(t)$ associated with a segment of trajectory of
length $t$ is equal, asymptotically, to
\begin{equation}
\lim_{|t| \to \infty} \frac{I(t)}{|t|} = h
% = \sum \lambda_+
\label{eq3} \end{equation}
where
%$\sum \lambda_+$ is the sum of all positive Lyapounov exponents
$h$ is the so called KS (Kolmogorov-Sinai) entropy which is positive
when the maximum Lyapunov exponent $\lambda$ is positive.\\
This means that in order to predict a new segment of a chaotic
trajectory one needs an additional information proportional to the
length of this segment and independent of the previous length of
the trajectory. In this situation information cannot be extracted
from the observation of past history of motion. This inherent
randomness exhibited by many Newtonian systems provides hope that
classical statistical mechanics can at last be rigorously derived
without the use of additional {\em ad hoc} assumptions such as
correlations decay and the like.

On the other hand one would like to know whether such a strong
property, being sufficient, is also necessary. Certainly
ergodicity only is not sufficient. Indeed it is known that
ergodicity alone, by merely implying that time averages are equal
to phase averages, is sufficient to justify equilibrium
statistical mechanics, but does not lead to any approach to
statistical equilibrium. For the latter, continuous spectrum of
the motion and correlations decay is necessary. This property, in
the ergodic theory of dynamical systems, is called {\em mixing}
and provides statistical independence of different parts of a
dynamical trajectory thus allowing a statistical description in
terms of few macroscopic variables. Mixing is therefore a stronger
property than ergodicity. However it does not require, in
principle, exponential instability. Recently strong empirical
evidence has been provided of mixing and diffusive behaviour in
dynamical systems with linear instability(zero Lyapunov
exponent)\cite{triang99,trmap00,trgas03,cond03,cond04}. In
addition, quite remarkably, normal heat transport, obeying the
Fourier law, has been found in a quasi-1d gas with triangular
scatterers \cite{trgas03} and in the 1d alternate masses hard
point gas \cite{cond04}. These results clearly demonstrate that
linear instability can be sufficient for the derivation of macroscopic
laws of physics.

Along these lines, a quite natural step forward is to analyze the
connection between linear dynamical instability and the statistical
entropy $S$. The latter, differently from the dynamical KS entropy,
is a function of time and depends on the initial probability
distribution (ensemble) for the state of the system. In
\cite{latora_1} it was numerically shown that, for symplectic
chaotic systems, the time evolution of the {\it phase-space
averaged} statistical entropy of out-of-equilibrium initial
ensembles displays a stage of linear increase with a slope that
coincides with $h$ (and, through the Pesin equality, with the sum of
positive Lyapunov exponents). It should be stressed that the above
numerical result, though very interesting, is far from being
rigorous. Actually, as pointed out also in a recent paper
\cite{vulpiani}, such connection does not hold in systems with
intermittent behavior or in systems which, though mixing and
exponentially unstable, are characterized by different time scales
in the relaxation process. This is the case of systems like the
Sinai billiard or the Bunimovich stadium in which the local Lyapunov
exponent is not constant.

In this paper we address precisely the above problem, namely the
connection between statistical entropy and dynamical instability,
for a system which is linearly (instead of exponentially)
unstable. It turns out that in order to recover results analogous
to those relating to exponentially unstable systems (see e.g.
\cite{latora_1}) a {\it generalization} of the definition of the
classical entropic functional $S_{BG}=-\sum_{i=1}^W p_i \ln p_i$
($BG$ stands for {\it Boltzmann-Gibbs}, $\sum_{i=1}^W p_i=1$) is
needed. A convenient such generalization has been proposed in 1988
\cite{tsallis} as  a basis for generalizing BG statistical
mechanics (for reviews see \cite{gellmann,swinney}). This new
functional is defined as
\begin{equation}
S_q=\frac{1-\sum_{i=1}^W p_i^q}{q-1} \;\;(q\in{\mathbb R},\;S_1=S_{BG}) \,.
\label{q_entropy}
\end{equation}
For equal probabilities $p_i$, we obtain $S_q=\ln_q W$, where
$\ln_q x \equiv (x^{1-q}-1)/(1-q) \;(\ln_1 x = \ln x)$. If we
extremize entropy (2) with appropriate constraints, we obtain
\cite{tsallis} $p \propto e_q^{-\beta_q {\cal H}}$ for the
canonical ensemble, where $\beta_q$ is an inverse-temperature-like
parameter, ${\cal H}$ is the Hamiltonian, and $e_q^x$ is the
inverse function of $\ln_q x$, called $q$-exponential, i.e.,
$e_q^x \equiv [1+(1-q)\,x]^{1/(1-q)} \; (e_1^x=e^x)$. For systems
for which $BG$ statistical mechanics is valid, we have $q=1$. For
anomalous systems (e.g., for long-range interactions \cite{gellmann,swinney,houches,EPNTamaritAnteneodo}), we expect
$q$ to be univocally determined by the (nonextensive) universality
class to which the system belongs. For example, for the edge of
chaos of unimodal maps it has been exactly shown \cite{robledo_1},
through a renormalization group approach, that: (i) In
correspondence with the Feigenbaum attractor
    the sensitivity to initial conditions has an upper-bound
    envelope which is algebraic in time and {\it precisely}
    given by a $q$-exponential function for the
    specific value $q=0.2445\ldots$ (this value is exactly
    deduced from 
    the $\alpha$ Feigenbaum's universal constant);
(ii) The connection between this power-law dynamical
    instability and a {\it linear} statistical entropy
    production-rate is obtained through the form
    (\ref{q_entropy}), for the same value of $q$.
These results are available through a number of different
methods (see \cite{gellmann,swinney} and references
therein for details).

In this paper we consider instead a conservative, linearly unstable
system, defined by the area preserving triangle map $z_{n+1}=T(z_n)$
\cite{trmap00} on a torus $z=(x,y)\in [-1,1)\times [-1,1)$
\begin{eqnarray}
y_{n+1} &=& y_n + \alpha {\rm\, sgn} x_n + \beta \pmod{2}, \nonumber \\
x_{n+1} &=& x_n + y_{n+1} \pmod{2},
\label{triangle}
\end{eqnarray}
where ${\rm sgn}x=\pm 1$ is the sign of $x$ and $\alpha,\beta$ are
two parameters ($n=0,1.\ldots$). The following facts and numerical
results have been established \cite{trmap00}: For rational values
of $\alpha,\beta$ the system is pseudo-integrable, as the dynamics
is confined on invariant curves. If $\beta=0$ and $\alpha$ is
irrational, the dynamics is ergodic but the phase-space is filled
very slowly (see also Ref.\cite{kaplan}), while for incommensurate
irrational values of $\alpha,\beta$ the dynamics is ergodic and
mixing with dynamical correlation function decaying as $t^{-3/2}$.
We would like to stress here that, as shown in \cite{trmap00}, the
triangle map does not have any secondary time scales and that the
exploration of the phase phase by a given orbit is arbitrarily
close to that of a random model.

\begin{figure}
\includegraphics[width=0.49\columnwidth]{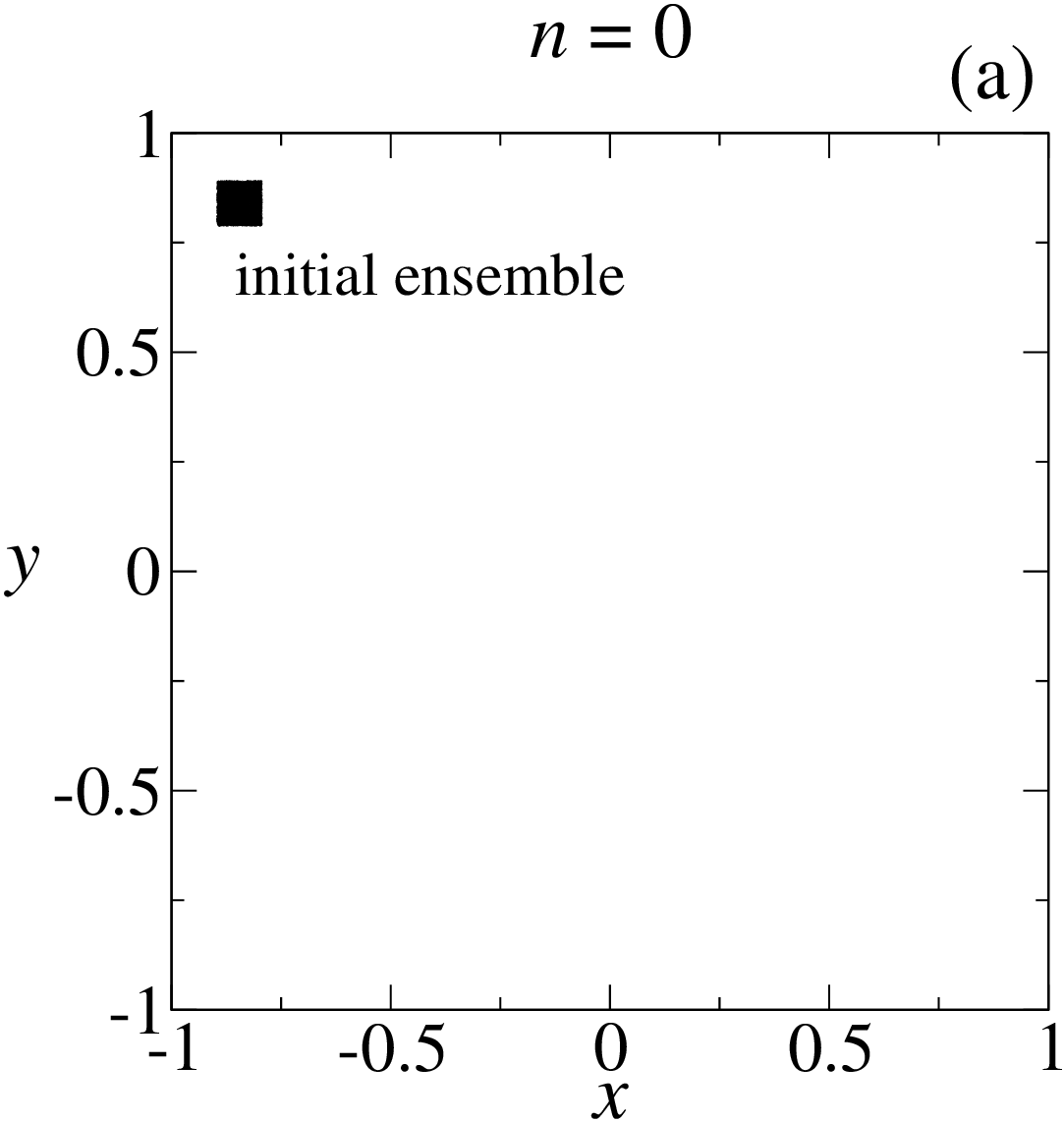}
\includegraphics[width=0.49\columnwidth]{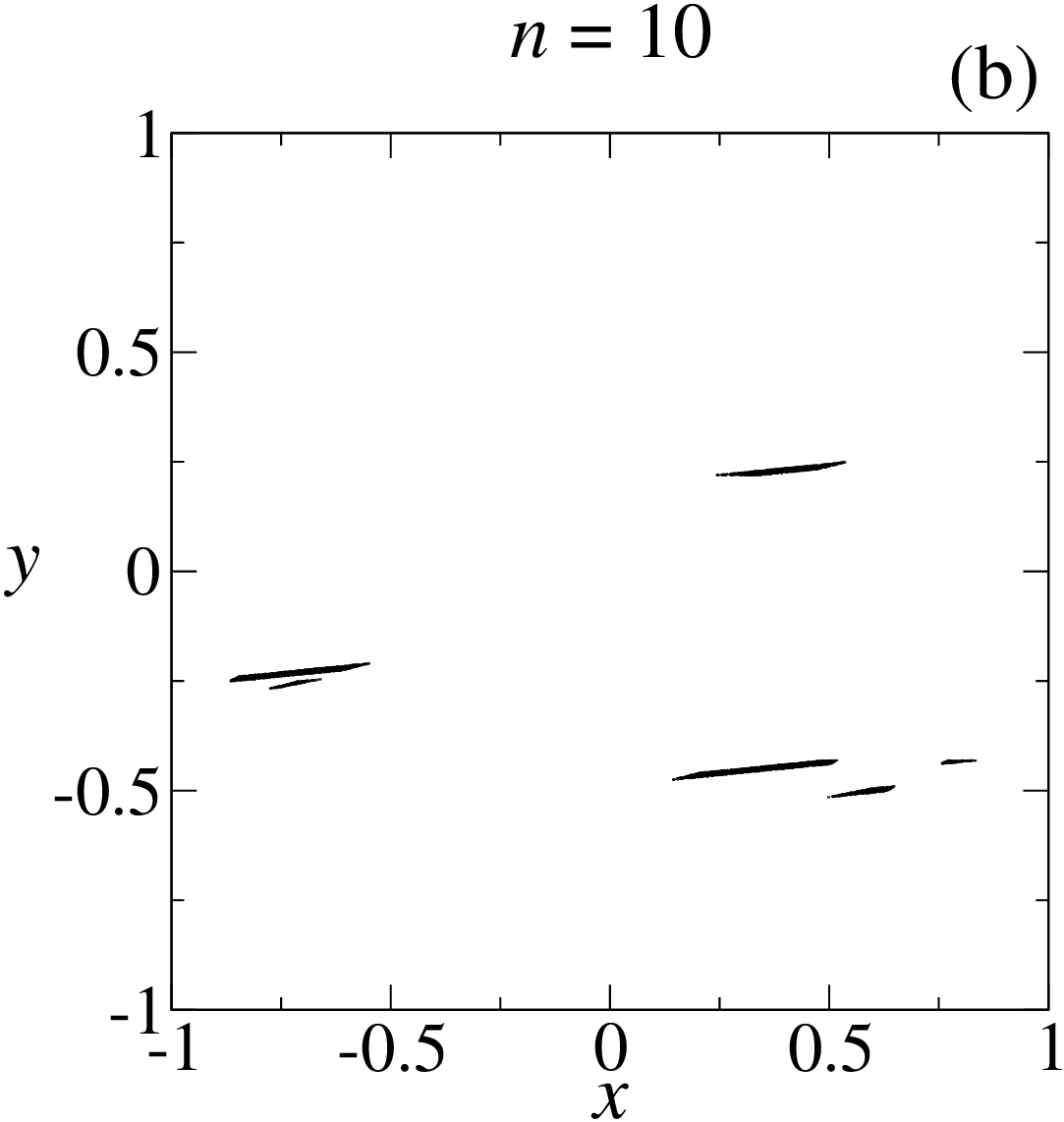}
\includegraphics[width=0.49\columnwidth]{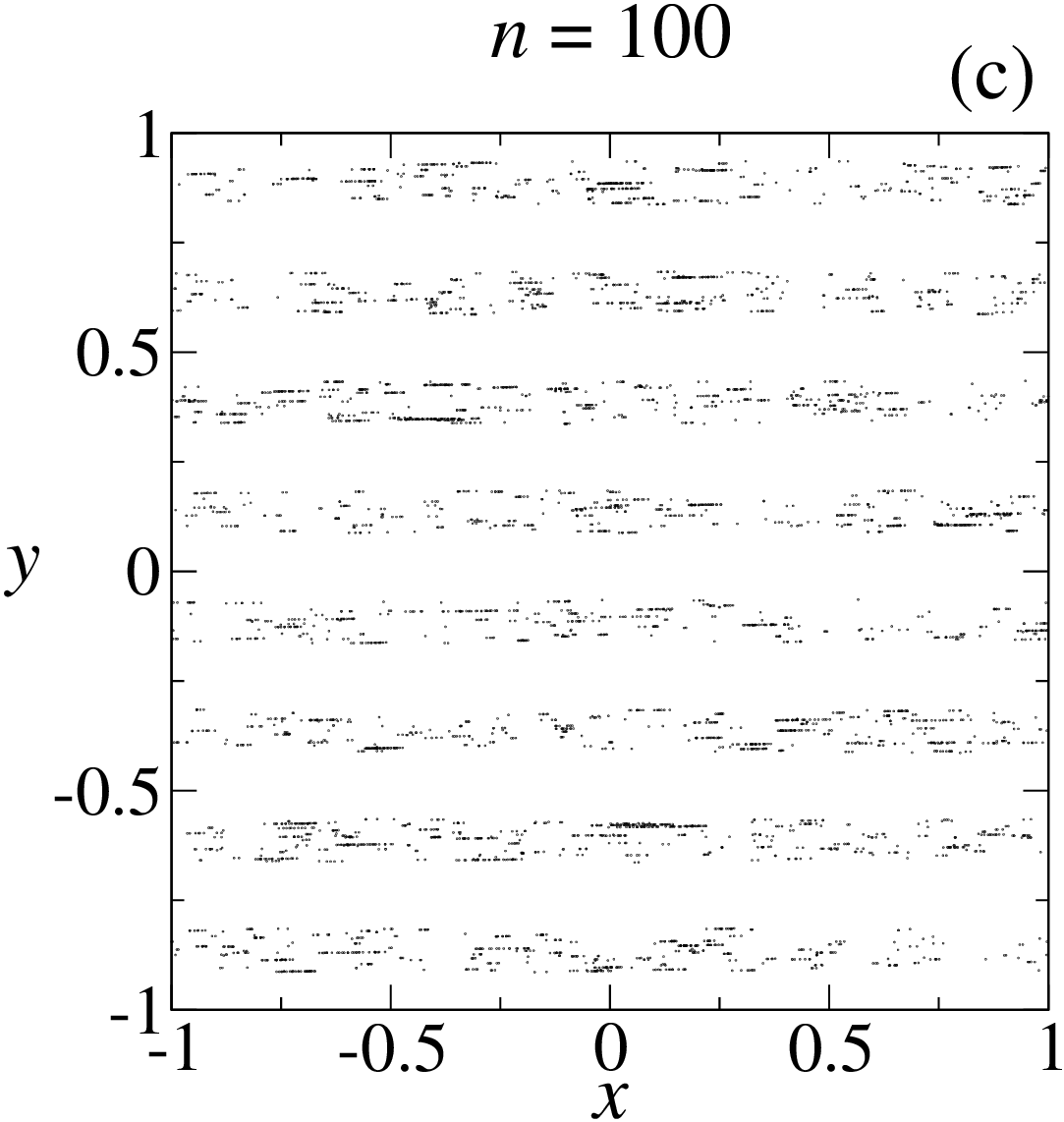}
\includegraphics[width=0.49\columnwidth]{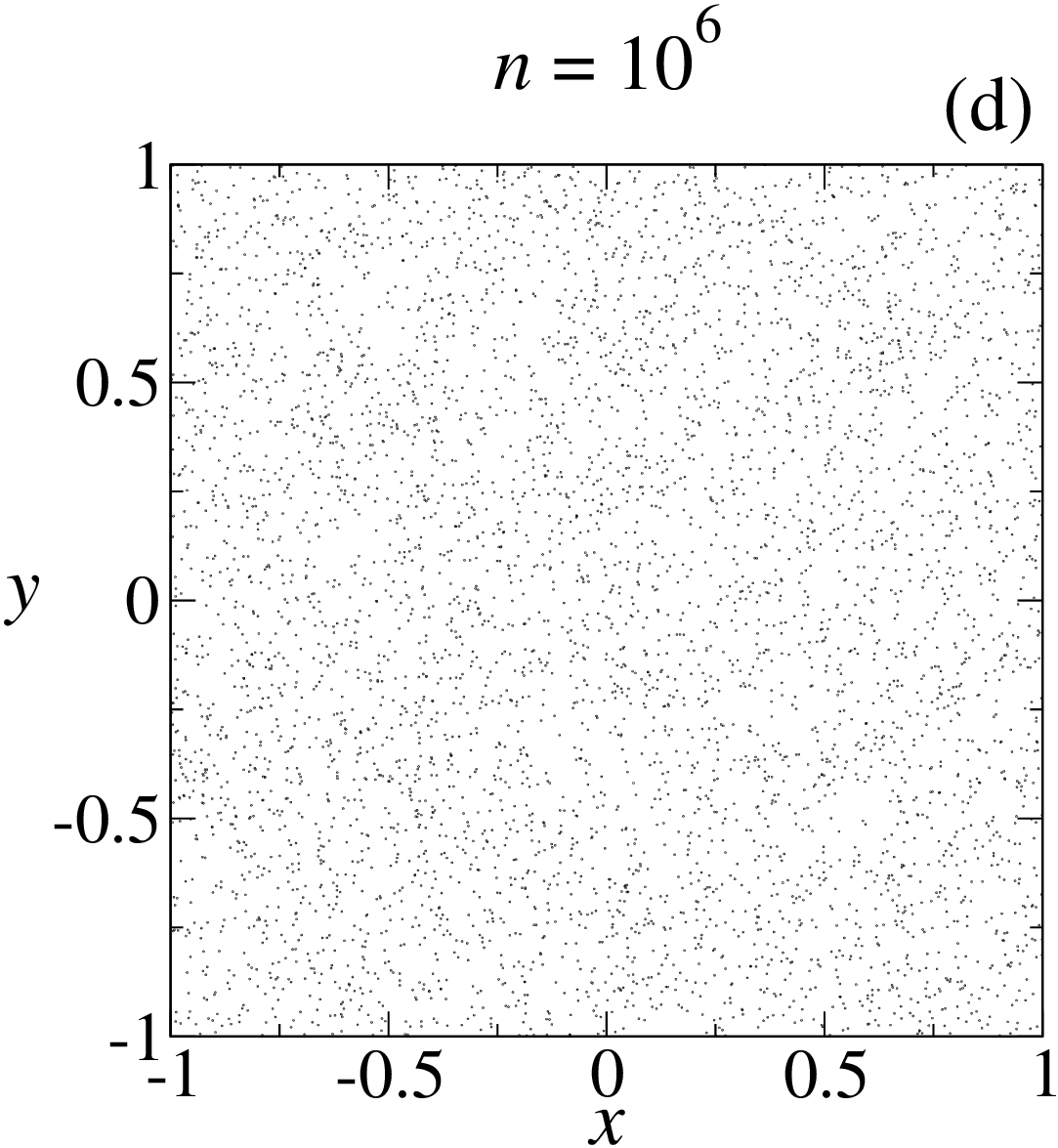}
\caption{Time evolution of an
  ensemble of points in phase space. (a) The ensemble is initially
  located inside a single cell. (b-d) Phase space distribution after $n= 10, 10^2,10^6$ map
  iterations.}
\label{fig_phase_space}
\end{figure}

It can be argued that the triangle map possesses essential
features of bounce maps of polygonal billiards and 1d hardpoint
gases \cite{triang99,trmap00}, namely the parabolic stability type
in combination with decaying dynamical correlations, and as such
represents a paradigmatic model for a larger class of systems. For
the sake of definiteness, in the following we will fix the
parameter values $\alpha=[\frac{1}{2}(\sqrt{5}-1)-e^{-1}]/2$,
$\beta= [\frac{1}{2}(\sqrt{5}-1)+e^{-1}]/2$ as it has been done in
\cite{trmap00}, although it should be noticed that qualitatively
identical results are obtained for other irrational parameter
values. Fig. \ref{fig_phase_space} shows the mixing process of an
ensemble of points initially localized inside a small square. The
action of the map (\ref{triangle}) initially divides the area
covered by the ensemble into different unconnected portions, each
essentially stretched along a straight line. After a certain
amount of time these portions overlap until a slow relaxation
process to a  complete mixing is observed.

In order to compute the dynamical evolution of the coarse grained
statistical entropy $S_q(n)$ we divide the phase space in $W\equiv
w\times w$ equal cells and consider an initial ensemble of $N$
points randomly distributed over a single partition-cell. Let
$p_i(n)$  the fraction of orbits which after $n$ time steps
overlap with the cell of label $i$. By means of Eq.
(\ref{q_entropy}) we can then compute $S_q(n)$. Typically, in
order to get reliable results the statistics needs to be improved
by averaging over many random initial cells \cite{latora_1}. For a
chaotic system with dynamical entropy $h$, one expects $S_1(n) = h
n$ in the limit $W\to\infty$  and for sufficiently large $n$
(i.e., after a possible initial transient) \cite{latora_1},
whereas for a finite $W$, $S_1(n)$ grows linearly with $n$ during
an {\it intermediate stage}, before saturation. In contrast with
such a behaviour, the time-evolution of the BG entropy associated
to the triangle map (\ref{triangle}) is characterized by an
intermediate stage where the entropy $S_1$ grows {\it
logarithmically} with time \cite{casati_log}.

Our numerical simulations of $S_q(n)$ provide clear empirical
evidence that, in analogy with the results for the edge of chaos
of unimodal maps \cite{robledo_1}, there is only a specific value
of $q\neq 1$ for which a linear entropy production rate is
observed when $W$ is sufficiently large. Indeed, only for $q=0$ we
get
\begin{equation}
S_0(n)=K\;n,
\end{equation}
with the numerical constant $K\simeq 1$. This production-rate
regime is associated to the first stage of the mixing process
(Fig. \ref{fig_phase_space}) which lasts infinitely long in the
limit $W\to\infty$. We remark that $S_0$ in Eq. (\ref{q_entropy})
reduces to $W(n)-1$, where $W(n)$ is the number of cells occupied
by the ensemble at the iteration step $n$. In Fig. 2(a)
we show the time evolution of $S_q$ for different values of the
parameter $q$. As the analysis of the derivative confirms (Fig. 2(b)), only for $q=0$ one gets a linear increase in
time. Fig. 2(c) shows that the linear increase with
slope $K\simeq 1$ is reached, in the limit $W\to\infty$, from
above. We remark that for a meaningful definition of the
probabilities $p_i$ the condition $N>>W(n)$ has to be fulfilled.

\begin{figure}
\begin{center}
\includegraphics[width=0.8\columnwidth]{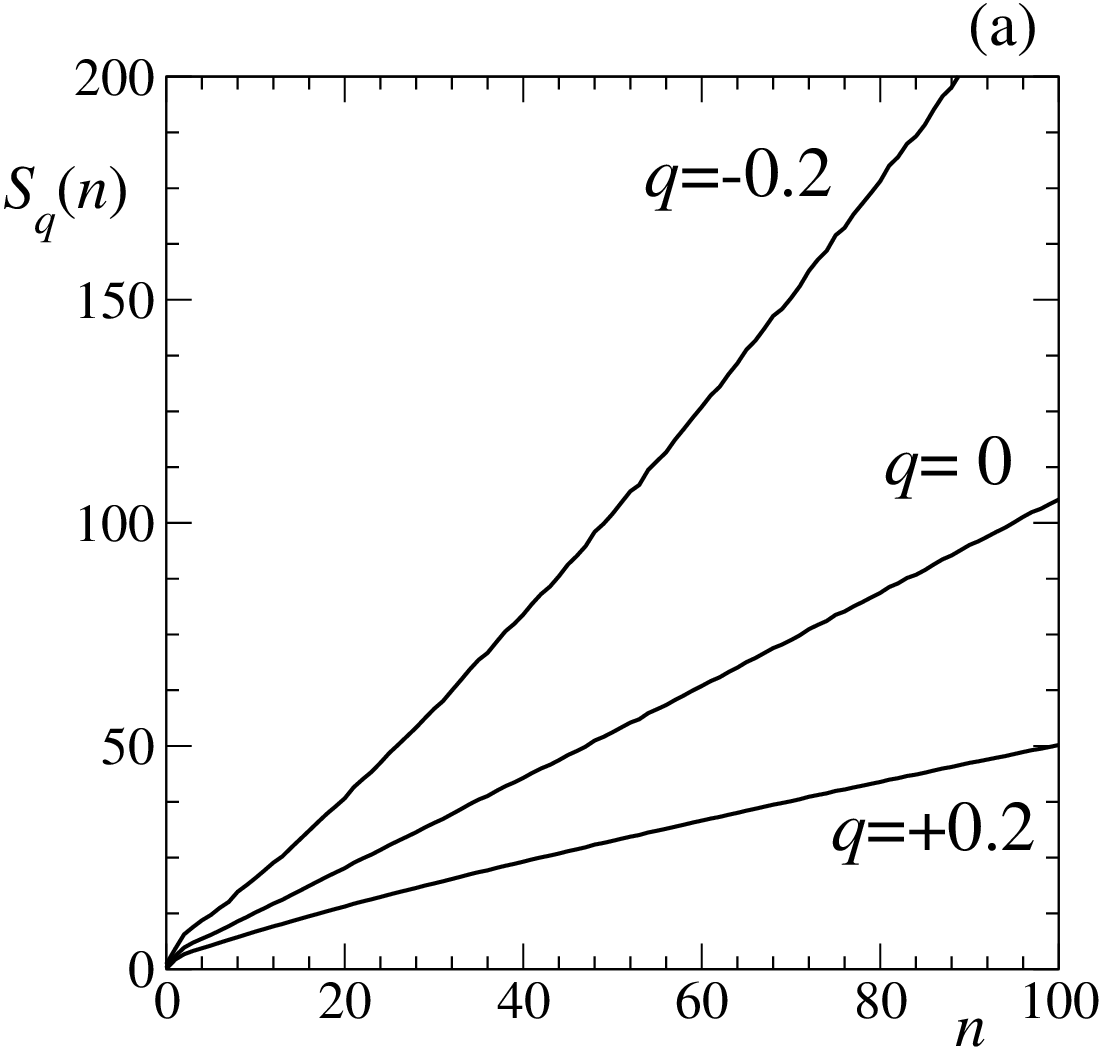}\\
\vspace{0.6cm}
\includegraphics[width=0.49\columnwidth]{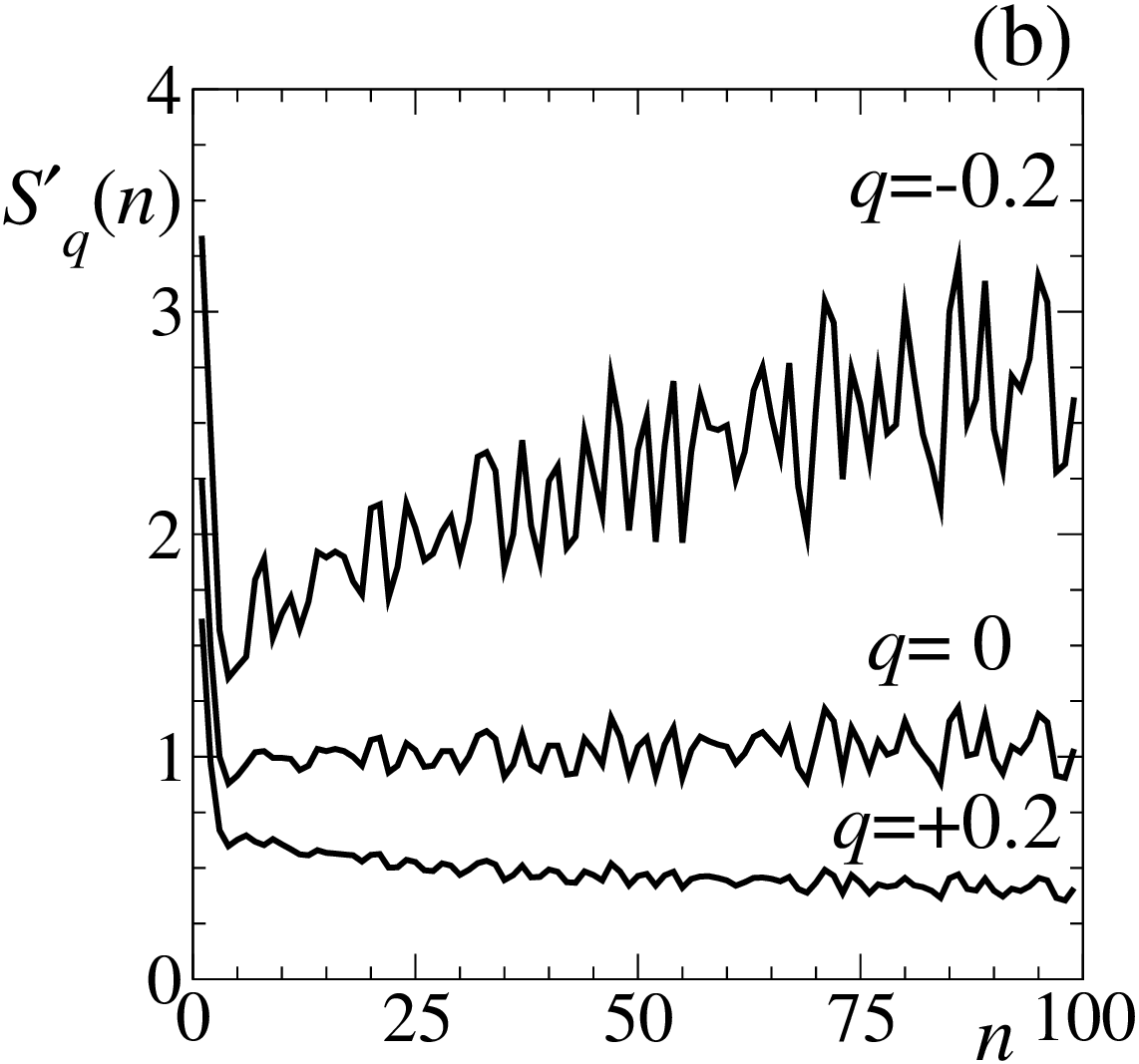}
\includegraphics[width=0.49\columnwidth]{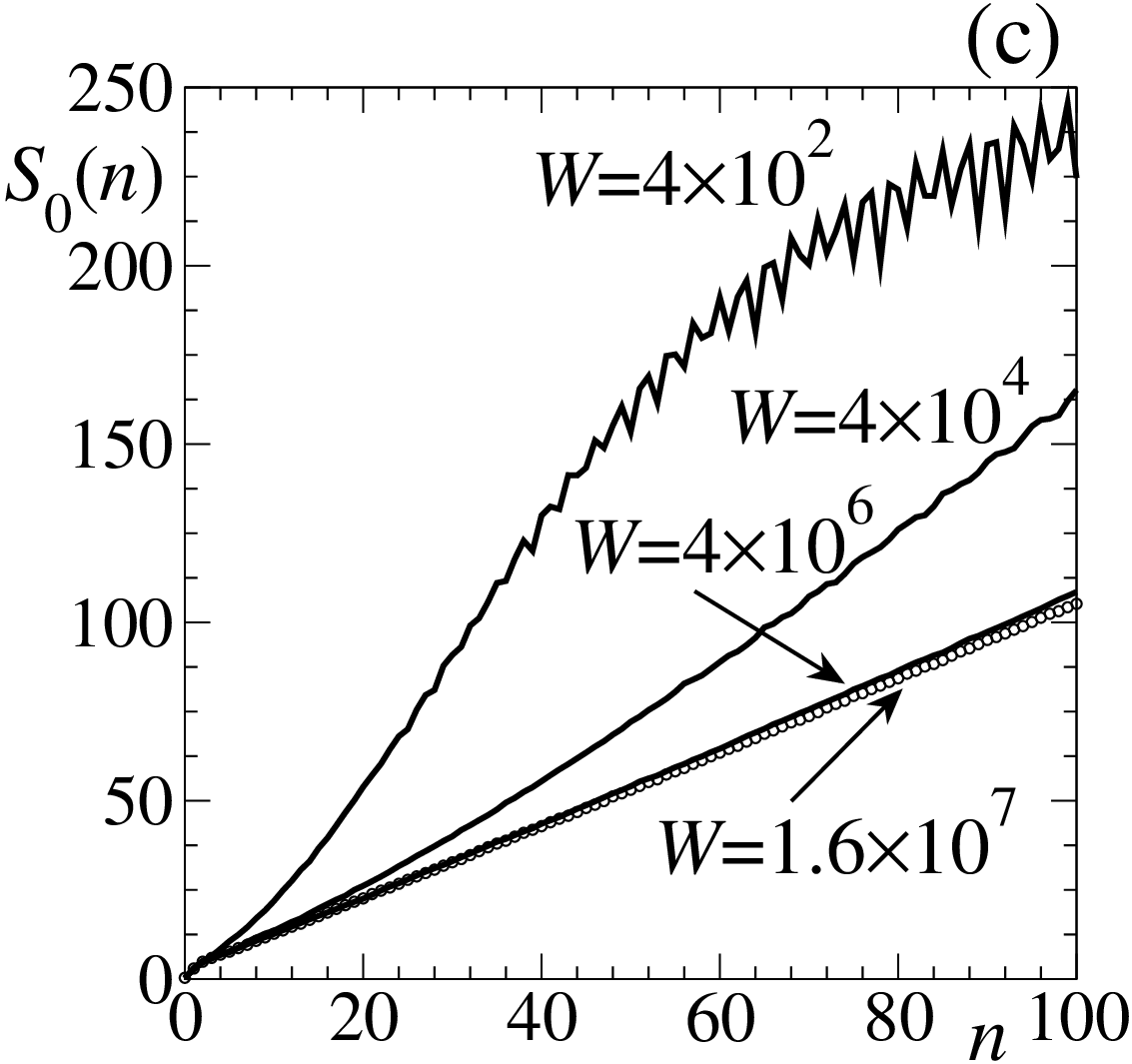}\\
\vspace{0.6cm}
\caption{Time-evolution of the statistical entropy
  $S_q$ for different values of $q$. The phase space has
  been divided into $W=4000\times4000$ equal cells of size
  $l=5\times10^{-4}$ and the initial ensemble is
  characterized by $N=10^3$ points randomly distributed
  inside a partition-square. Curves are the result of an
  average over $100$ different initial squares randomly
  chosen in phase space. The analysis of the derivative of
  $S_q$ in (b) shows that only for $q=0$ a
  linear behaviour is obtained. In fact, a linear regression
  provides $S_0(n)=1.029\;n+1.997$ with a correlation
  coefficient $R=0.99993$. (c) shows that the linear growth
  for $S_0$ is reached from above, in the limit $W\to\infty$.}
\end{center}
\label{fig_s_q}
\end{figure}

The above numerical result directly follows from the fact that the
mixing process occurs essentially along straight lines. Indeed we
have \footnote{The correction factor due to the orientation of the
stretched ensemble with respect to the partition grid becomes
negligible as $|\Delta {\mathbf x}(n)|>>l$. }
\begin{equation}
W(n)\simeq\frac{|\Delta {\mathbf x}(n)|}{l},
\label{W_sen}
\end{equation}
where $|\Delta {\mathbf x}(n)|$ is the length of the segment that
describes the maximum separation between points of the ensemble at
time $n$, and $l$ is the size of a single cell partition. If we
linearize map (\ref{triangle}) then the growth of the tangent
vector is given by
\begin{eqnarray}
\Delta y_{n+1} &=& \Delta y_n \pmod{2}, \nonumber \\
\Delta x_{n+1} &=& \Delta x_n + \Delta y_{n} \pmod{2},
\label{triangle_linearized}
\end{eqnarray}
which leads to
\begin{equation}
|\Delta {\mathbf x}(n)|\sim n|\Delta y_0|. \label{delta_x_n}
\end{equation}
This linear increase of $|\Delta {\mathbf x}(n)|$ implies
the linear increase of $W(n)$ in (\ref{W_sen}).
Notice that if the
partition is not fine enough,  relation (\ref{W_sen}) breaks down.
This reflects in a faster growth of $S_0$ for
small values of $W$, as observed in Fig. \ref{fig_s_q}c.

Next, we discuss the connection between the statistical
entropy and the sensitivity to initial
conditions defined as
\begin{equation}
\xi(n)\equiv\frac{|\Delta {\mathbf x}(n)|}
{|\Delta {\mathbf x}(0)|}.
\end{equation}
From the linearized map (\ref{triangle_linearized}) we see that
the sensitivity to initial conditions for the triangle map casts
into the $q$-exponential form provided by the afore-mentioned
generalization of the BG formalism (see e.g.
\cite{gellmann,swinney}): $ \xi =[1+(1-q)\lambda_q n]^{1/(1-q)}$,
with $q=0$ and $\lambda_0=|\sin\theta_0|$, $\theta_0$ being the
polar angle in the $(x,y)$-plane of the initial vector ($|\Delta
y_0|=|\sin\theta_0||\Delta {\mathbf x}(0)|$). Taking the
upper-bound of the sensitivity to initial conditions we get

\begin{equation}
K=\max\{\lambda_0\}=1.
\label{lambda_k}
\end{equation}

Eq. (\ref{lambda_k}) is in some sense a generalization of the
Pesin equality to a system with zero Lyapunov exponent, in fact it
relates the dynamical instability to an entropic quantity. In
spite of its simplicity this is a non trivial result(analogous to
that exhibited in \cite{robledo_1}) If, on the other hand, one
considers the average value of $\xi(t)$ over an ensemble of
uniformly distributed initial directions $\theta_0$ (due to
ergodicity  this coincides with the long-time-average), one has
$K=(\pi/2)\langle\lambda_0\rangle$, since
$\langle\lambda_0\rangle=\langle|\sin \theta_0|\rangle=2/\pi$.

In conclusion, while positivity of Lyapunov exponents is {\it
sufficient} for a meaningful statistical description (the $BG$
statistical mechanics), it might be {\it not necessary}.
 Indeed, we have illustrated, for a conservative, mixing and ergodic nonlinear dynamical system,
 that the use of the more general entropy $S_q$ (with the value
$q=0$ for this case)
  provides a satisfactory frame for handling nonlinear dynamical systems whose
   maximal Lyapunov exponent vanishes. In particular,
   we have shown that  (the upper bound of) the coefficient $\lambda_q$ of the
   sensitivity to the initial conditions {\it coincides} with the entropy production
   per unit time, in total analogy with the Pesin theorem for standard chaotic systems.
These results suggest that a thermostatistical approach of such systems is possible.
Indeed, the structure that we have exhibited here for the time dependence of $S_q$,
is totally analogous to the one that has been recently exhibited \cite{TGS} for
the $N$-dependence of $S_q$ , where $N$ is the number of elements of a many-body system.
 When the number of nonzero-probability states of the system increases as a power
 of $N$ (instead of exponentially with $N$ as usually), a special value of $q$ below
 unity exists such that $S_q$ is {\it extensive}. In other words, $S_q$ asymptotically
 increases {\it linearly} with $N$, whereas $S_{BG}$ does not.

   In this paper we have considered an abstract model of dynamical
   systems. However the same features of  linear instability
   together with a nice diffusive behaviour are exhibited by more
   realistic systems like billiards in irrational polygons and one
   dimensional systems of unequal masses hard point particles.
   On one hand this fact generalizes the validity of our
   conclusions. On the other hand all these systems belong to the
   extreme case of linear instability leading to the value $q =0$.
   Certainly it would be much more interesting to find dynamical
   systems leading to $  0< q <1 $. Natural candidates are systems with power law
   instability $t^\alpha$ with $\alpha >1$. We are not aware of
   any conservative system with such property and with nice mixing behaviour. It would be highly
   interesting  to explore such possibility.

\section{Acknowledgments}
GC is partially supported by EU Contract No.
HPRN-CT-2000-0156(QTRANS) . We have also benefited from partial
support by CNPq, Faperj and Pronex (Brazil), and by SI International and Air Force Research Laboratory (USA).

%\bigskip
%\newpage

\end{document}